# Which percentile-based approach should be preferred for calculating normalized citation impact values? An empirical comparison of five approaches including a newly developed citation-rank approach (P100)

Lutz Bornmann,[#] Loet Leydesdorff,[*] and Jian Wang[§+]

[#] Division for Science and Innovation Studies

Administrative Headquarters of the Max Planck Society

Hofgartenstr. 8,

80539 Munich, Germany.

E-mail: bornmann@gv.mpg.de

[*] Amsterdam School of Communication Research (ASCoR),

University of Amsterdam, Kloveniersburgwal 48,

1012 CX Amsterdam, The Netherlands.

Email: loet@leydesdorff.net

[§] Institute for Research Information and Quality Assurance (iFQ)

Schützenstraße 6a, 10117 Berlin, Germany.

[+] Center for R&D Monitoring (ECOOM) and Department of Managerial Economics, Strategy and Innovation, Katholieke Universiteit Leuven, Waaistraat 6, 3000 Leuven, Belgium.

Email: Jian.Wang@kuleuven.be

**Abstract**

For comparisons of citation impacts across fields and over time, bibliometricians normalize the observed citation counts with reference to an expected citation value. Percentile-based approaches have been proposed as a non-parametric alternative to parametric central-tendency statistics. Percentiles are based on an ordered set of citation counts in a reference set, whereby the fraction of papers at or below the citation counts of a focal paper is used as an indicator for its relative citation impact in the set. In this study, we pursue two related objectives: (1) although different percentile-based approaches have been developed, an approach is hitherto missing that satisfies a number of criteria such as scaling of the percentile ranks from zero (all other papers perform better) to 100 (all other papers perform worse), and solving the problem with tied citation ranks unambiguously. We introduce a new citation-rank approach having these properties, namely P100. (2) We compare the reliability of P100 empirically with other percentile-based approaches, such as the approaches developed by the SCImago group, the Centre for Science and Technology Studies (CWTS), and Thomson Reuters (InCites), using all papers published in 1980 in Thomson Reuters Web of Science (WoS). How accurately can the different approaches predict the long-term citation impact in 2010 (in year 31) using citation impact measured in previous time windows (years 1 to 30)? The comparison of the approaches shows that the method used by InCites overestimates citation impact (because of using the highest percentile rank when papers are assigned to more than a single subject category) whereas the SCImago indicator shows higher power in predicting the long-term citation impact on the basis of citation rates in early years. Since the results show a disadvantage in this predictive ability for P100 against the other approaches, there is still room for further improvements.

# 1 Introduction

For comparisons of citation impacts across fields and over time, bibliometricians normalize the observed citation counts with reference to an expected citation value. For many years, this expected citation value was commonly calculated as the (arithmetic) average citations of the papers in a reference set; for example, the citation counts of all papers published in the same subject category and year as the paper(s) under study can be averaged. However, the arithmetic average of a citation distribution hardly provides an appropriate baseline for comparison because these distributions are extremely skewed (Seglen, 1992): The average is heavily affected by a few highly cited papers (Waltman et al., 2012).

Percentile-based approaches (quantiles, percentiles, percentile ranks, or percentile rank classes) have been proposed as non-parametric alternative to these parametric central-tendency statistics. Percentiles are based on an ordered set of citation counts (i.e., papers are sorted in ascending order of citation counts), whereby the fraction of papers at or below the citation counts of a focal paper is used as an indicator for the relative citation impact of this focal paper. Instead of using an average citation impact for normalizing a paper under study, its citation impact is evaluated by its rank in the citation distribution of similar papers in a reference set (Leydesdorff & Bornmann, 2011; Pudovkin & Garfield, 2009).

This percentile-based approach arose from a debate in which we argued that frequently used citation impact indicators based on using arithmetic averages for the normalization— e.g., "relative citation rates" (Glänzel, Thijs, Schubert, & Debackere, 2009; Schubert & Braun, 1986) and "crown indicators"(Moed, De Bruin, & Van Leeuwen, 1995; van Raan, van Leeuwen, Visser, van Eck, & Waltman, 2010)—had been both technically (Lundberg, 2007; Opthof & Leydesdorff, 2010) and conceptually (Bornmann & Mutz, 2011) flawed. The non-parametric statistics for testing observed versus expected citation distributions were further elaborated by Leydesdorff, Bornmann, Mutz, and Opthof (2011). Various statistical procedures have been proposed to analyze percentile citations for institutional publication sets (Bornmann, 2013; Bornmann & Marx, in press; Bornmann & Williams, in press).

In this study, we pursue two related objectives—an analytical and an empirical one: (1) Although different percentile-based approaches have been developed (see an overview in Waltman & Schreiber, 2013), an approach is missing that satisfies, in our opinion, a number of important criteria such as scaling from the percentile rank 0 (all other papers perform better) to 100 (all other papers perform worse), solving the problem with tied citation ranks (papers with the same number of citations in a reference set) unambiguously, ensuring that the mean percentile value is 50, and asymmetrically handling the tails of the distributions. Bibliometricians hitherto use different approaches, but unfortunately no approach for the computation of percentiles in the case of discrete distributions is without disadvantages (Hyndman & Fan, 1996).

In the following, we propose a new citation-rank approach which, in our opinion, solves the two major problems analytically: (i) unambiguous scaling from 0 to 100, and (ii) equal values for tied ranks. In order to achieve this objective, we use a distribution other than the citation distributions, namely the distribution of *unique* citation values. On this basis, papers with tied citations obtain necessarily the same rank. Furthermore, it seems to us that the scale can run from 0 to 100 with the highest-ranking paper in a reference set at 100 and the lowest at 0. In other words, we derive the ranks from the empirical distribution of the unique citation counts in the reference set. By definition, the highest-ranking papers in different reference sets all obtain the highest rank of 100, the lowest ones are fixed at zero, and thus, the distributions are made comparable. By defining the citation ranks empirically as intervals between the two extremes (of 0 and 100), one gains a degree of freedom which enables us to solve the problem of the otherwise floating maximum or minimum values of percentiles in discrete distributions.

(2) We propose to use the abbreviation "P100" for our new approach. In the second part of this study we compare the reliability of our new citation-rank approach with percentile-based approaches, among which are the approaches developed by the SCImago group for the ranking of universities and research-focused institutions and by the Centre for Science and Technology Studies (CWTS) for the ranking of universities. Using all papers in Web of Science (WoS, Thomson Reuters) published in 1980 (Wang, 2013), we retrieve for each subsequent year the citation counts of all these papers. The citation impact is then calculated based on the different percentile-based approaches and P100. The results show differences among the approaches in assigning the papers to various rank classes (e.g., the 10% most frequently cited papers in a certain subject category) and in the ability of estimating the long-term citation impact of the papers (their cumulative citation impact in the final [31$^{st}$] year (2010). We are most interested in the degree of agreement between the cumulative citation impact in the final year and the impact in early years (especially the first few years after publication).

There is no reason to assume that the first three to five years of citation can be used as a predictor of the long-term citation impact of papers. Wang (2013) found low correlations between citation counts in short time windows and total citations counts in 31 years, and these correlations are even lower for field-normalized citation counts and highly cited papers. By applying Group-Based Trajectory Modeling (Nagin, 2005) to citation distributions of journals, Baumgartner and Leydesdorff (in press) showed that papers within an otherwise similar set can vary significantly in terms of how they are cited in the longer run of ten or fifteen years.

Given the long time-window of 31 years in this study, one can assume that we capture the entire citation impact in the long run (in other words, with the long run we will have a valid estimate of a paper's "true" citation impact). The analytical argument about the rules for defining percentiles or citation ranks, however, is a different one from the empirical usability of one or another (percentile-based) approach as a predictor in evaluation studies. The latter is a correlation which does not imply a causal explanation for the preference of the one or the other approach.

## 2 Methods

### 2.1 Different percentile-based approaches for normalizing citation counts

Three steps are needed in order to calculate percentiles for a reference set:

First, all papers in the set are ranked in ascending order of their numbers of citations. Papers with equal citation counts are often set equal by assigning the average rank. For example, this is the default ranking method in the statistical package Stata (StataCorp., 2011). This method ensures that the sum of the ranks is fixed at $n*(n+1)/2$, where $n$ is the number of papers in a reference set.

Second, each paper is assigned a percentile based on its rank (percentile rank). Percentiles can be calculated in different ways (Bornmann, Leydesdorff, & Mutz, 2013; Cox, 2005; Hyndman & Fan, 1996). The most commonly used formula is $(100*(i-1)/n)$, where $n$ is the total number of papers, and $i$ the rank number in ascending order. For example, the median value or the 50$^{th}$ percentile rank separates the top-half of the papers from the lower half. However, one can also calculate percentiles as $(100*(i/n))$. This calculation is used for the percentile data provided in InCites[1] and, for example, by Rousseau (2012). (Note that InCites sorts citations in descending—instead of ascending—order and then uses the maximum rank for tied papers. We shall use in this case the complement value of 100-$x$ for the attribution.)

One often uses a compromising formula, $(100*(i-0.5)/n)$, derived by Hazen (1914, p. 1549f.). This formula is used very frequently nowadays for the calculation of percentiles and is wired into the official Stata command "quantile" (StataCorp., 2011). It ensures that the mean percentile is 50 and symmetrically handles the tails of the distributions. Leydesdorff (2012) specified the uncertainty in the percentile attribution as ($0.5*n$). On the basis of ($i/n$), this correction leads to (($i/n$)-($0.5/n$)=($i$-0.5)/$n$) in the case of using Rousseau's (2012) formula for the computation or ((($i$-1)/$n$)+($0.5/n$)=($i$-0.5)/$n$) using the formula of Leydesdorff, et al. (2011). In both cases, the result is precisely the same as Hazen's (1914) compromise value.

---

[1] InCites (Thomson Reuters, http://incites.thomsonreuters.com/) is a web-based research evaluation tool allowing the evaluation of the productivity and citation impact of institutions.

Third, the minimum or maximum of the percentile rank can be adjusted. Papers with zero citations can be assigned a rank of zero. Only by assigning the rank zero to the papers with zero citations, one ensures that the missing citation impact of these papers is reflected in the percentiles in the same way in every case. Different ranks for papers with zero citations would arise if percentiles are calculated without using a constant rank of zero at the bottom (Leydesdorff & Bornmann, 2012; Zhou & Zhong, 2012).

A technical issue in the case of using percentiles for research evaluation pertains to the handling of ties (e.g., Pudovkin & Garfield, 2009; Rousseau, 2012; Schreiber, 2013, in press; Waltman & Schreiber, 2013). Which percentile ranks should be attributed to papers in a reference set with the same numbers of citations? Schreiber (2012) solved this problem by using the average of the percentile ranks, and Waltman and Schreiber (2013) further proposed fractional counting in order to attribute the set under study to percentile rank classes that are pre-defined, for example, for administrative reasons (e.g., for the purpose of comparative evaluations of universities).

For example, if one wishes to use the top 10% most frequently cited papers as an excellence indicator in university rankings (Waltman, et al., 2012), the tying of the ranks at this threshold level can generate an uncertainty. By proportional attribution of the fractions to the different sides of the threshold, this uncertainty can be removed from the resulting indicator. However, this approach can only be used to determine the exact proportion of the top 10% (or $x$%) papers in a reference set, but fractional counting cannot be used for the calculation of percentile ranks of the individual papers under study. In most studies, individual papers are the units of analysis. Following Hazen (1914), Leydesdorff (2012) suggested that the specification of the uncertainty at this lower level enables us to use statistics. In summary, the fractional attribution of percentile ranks is computationally intensive and for many research questions not functional.

SCImago—another provider of university and journal rankings (Bornmann, de Moya Anegón, & Leydesdorff, 2012)—uses another solution for the problem of ranks tying at the 10% threshold level. This group introduces a secondary sort key in addition to citation counts: When the citation counts are equal, the publication in a journal with the higher SCImago Journal Rank (SJR2) (Guerrero-Bote & de Moya-Anegon, 2012) obtains the higher percentile rank. Adding this journal metric takes into account not only the observed citations of the focal paper but also the prestige of the journal in which this focal paper is published.

A disadvantage of the SCImago approach, in our opinion, is the lack of a rationale for using a journal indicator (e.g., SJR2) to rank individual papers with the same citation counts. Why should a paper published in a higher-ranked journal receive a higher rank at the paper level than a paper published in a less reputable journal when both papers gathered the same number of citations? One could argue also the other way around: The paper in the less reputable journal could receive the higher rank, although it was not supported by the boost of publishing in a more reputable journal (e.g., *Nature*). From this perspective, the latter paper performed better with an equal citation score than the paper in the more highly-ranked journal.

## 2.2   P100: A new citation-rank approach

Our new approach is motivated by the following two desirable properties: (1) In the case of ties in the citation distribution, papers with the same number of citations should be attributed to the same rank, and fractional counting should be avoided for the reasons specified above (i.e., to have paper-based ranks for the statistical analyses). (2) In each reference set, the rank classes should be exactly distributed from 0 to 100. Working with fixed values for the papers with the highest and lowest citation counts solves the problem in previous percentile-based approaches that the minimum or the maximum percentile rank can vary across different reference sets.

For example, if 50% of the papers in a reference set remained un-cited, Rousseau's (2012) formula would assign all these papers to the 50$^{th}$ percentile. Using Hazen's (1914) formula, high-impact papers equally ranked at the top of the distribution do not necessarily receive a percentage of 100; these problems especially arise in the case of relatively small reference sets (such as for journal sets including reviews, Leydesdorff & Bornmann, 2011). In our opinion, it is inappropriate in an evaluation study that the most (or least) highly cited paper in one reference set receives a different percentile than the most (or least) highly cited paper in another reference set because of unsolved problems with and disagreements about the normalization. When both papers have the same top performance (there is no other paper with a better performance), they should both receive the same rank of 100. The same is true for the least-cited paper in a reference set.

In our new approach the highest citation count in the reference set receives always a rank of 100 and the lowest a rank of 0. The ranks are subsequently calculated as $(100 / i_{max}) * i$, whereby the papers in the reference set are sorted from 0 to $i_{max}$ (in increasing order of the unique citation counts). Let us calculate the ranks for an example of seven papers in a fictitious reference set in Table 1. The two grey shaded papers are the papers under study for which reference values are needed: (1) The first paper has a rank of 66.4 since a paper with 4 citations is attributed to this rank. (2) The second paper receives a rank of zero. Using the above mentioned formula of Rousseau (2012), that is $((i/n)*100)$, however, this paper would have been attributed to the rank of 14.3. The lowest (and highest) values would thus become variable in different reference sets using percentile-based approaches. P100 is no longer percentile-based, but citation-values based.

Concerning the problem of tied ranks among citations, P100 starts with the analytical premise that the distribution of citations over papers is not the interesting one for the reference set, but the distribution of the unique citation values. The number of occurrences of these values in samples can be considered as an empirical effect which can analytically be neglected for the normalization. First, one would like to know which citation impact is the highest and which is the lowest in a reference set. We propose to rank subsequently the "unique" citation values instead of the citation distributions over all papers in a reference set. Papers with the same citation counts in the reference set are considered *only once in this distribution*. Thus, the distribution of the papers (with different or the same citation counts) is reduced to the distribution of unique citation counts (as against the frequency distribution of these citation counts).

To illustrate the proposed procedure, Table 2 shows another fictitious reference set which is used again for two papers under study. The paper with 38 citations receives a rank of 57.2 and the three papers with four citations a rank of 14.3. The reference classes are defined by ((1/7)*100). They are equally divided among the eight values of the unique citation counts.

In summary, the advantages of P100 can be listed as follows:

- The paper with the highest impact always receives the same rank of 100 (independently of the distribution of citation counts in different reference sets). The same is true for the paper with the lowest citation impact, receiving the value of 0.
- If there is more than one paper in a reference set with the same citation count, they are attributed to the same rank.
- The use of the distribution of unique citations instead of the citation counts of all papers in a reference set for the calculation of citation ranks solves the problems using different rules (proposed up to now), and avoids complicated methods for dealing with ties, such as the fractional approach (Schreiber, 2013; Waltman & Schreiber, 2013).

## 2.3 Dataset used for the empirical comparison of the different percentile-based approaches and P100

The initial dataset contains all journal papers published in 1980 and indexed in WoS, that is, 746,460 papers in total. Two restrictions are then imposed on the sample: (1) three document types – articles, reviews, and notes [2] – were kept while other documents types – that is, 119,797 papers – were excluded, and (2) only papers having at least one reference set with hundred or more papers, were included. (This restriction will be further explained in the following two paragraphs.)

In this study, we define a reference set for a paper under study as a set of papers with the same WoS subject category and document type. We have 713 reference groups (unique subject category and document type combinations): 247 article reference groups, 237 note reference groups, and 229 review reference groups with corresponding unique subject categories (247 unique categories). Because of the second restriction of 100 documents in a reference set, we have 228 article reference groups, 132 note reference groups, and 22 review reference groups (228 unique subject categories) in the data set. Although we have disproportionately excluded more reviews, this is very unlikely to cause a significant bias in our results. Wang (2013) showed that articles, notes, and reviews have more similar citation ageing patterns, compared with other most frequent document types – letters, meeting abstracts, and editorial materials. Taken as a whole, we only excluded 7.7% (n=19) subject categories (at the subject category level) and 24.1% (151,272) papers (at the paper level). As a result, we kept 475,391 papers for the analysis, and the annual citation counts (from 1980 to 2010) of these papers were retrieved from WoS.

---
[2] Notes were removed from the database as a document type in 1997, but they were citable items in 1980.

The reference sets were used to calculate values of P100 and the following four percentile-based approaches: the approaches developed by Hazen, Thomson Reuters (InCites), SCImago, and CWTS. Each paper in WoS is classified into one unique document type but possibly into multiple subject categories. Therefore, for papers with multiple subject categories, different aggregation rules were implemented to construct a unique value (rank) for each paper. The average (percentile) rank is used for the approaches of Hazen, SCImago, and P100, and the best performance across reference sets is used for the approach used for InCites.

When using the CWTS approach based on fractionation, the average probability of belonging to a specific group is considered as a fraction. Specifically, for any pre-specified (percentile) rank class (e.g., the top $x$% highly cited papers), each paper gets a number $p$ indicating its probability of belonging to this top $x$% class ($0 \leq p \leq 1$). However, a paper can get different values for $p$ in multiple reference sets, and the average value of $p$ is used to construct a unique indicator for this paper (Ludo Waltman, personal communication, 5/30/2013).

Furthermore, the citation impact values (percentiles or ranks) could be too discrete if the size of the reference set becomes too small. Therefore, only reference sets with at least hundred papers are included.[3] For example, if a paper belongs to two different reference sets: *A* and *B*, and *A* has more than 100 papers while *B* has less than 100 papers, then the impact values based on *B* are discarded. If neither *A* nor *B* has more than 99 papers, then both results based on *A* and *B* are discarded, and this paper is excluded from the further analysis. As the reference sets are defined not only by subject category, but also by document type, it is likely that the number of subject categories for reviews and notes is reduced more significantly than for articles. Reviews and notes are less frequent than articles and therefore are more likely to be excluded by the restriction of requiring at least 100 papers in the reference set.

---

[3] We decided to use 100 papers as a limit to produce reliable data. There is a high probability that the use of a limit of 50 or 200 would come to similar results as ours.

## 3 Results

For the investigation of the different percentile-based approaches and P100, we undertake three empirical analyses:

(1) First, we compare the distribution of impact values which are calculated by the different approaches (Hazen, InCites, SCImago, and P100) on the base of a fictitious reference set of 50 papers.

(2) In the second analysis we follow the annual citation impacts of the papers from 1980 up to 31 years later (that is, up to 2010). We are interested in how consistently the different approaches (Hazen, InCites, CWTS, SCImago, and P100) assign these papers to different rank classes (top 50%, top 10%, top5%, and top 1%) across the years and whether different approaches behave similarly or one approach outperforms the others.

(3) The third part of the empirical analysis focuses on a comparison of the practical use of the different approaches. Based on the papers published in 1980, we randomly sampled fictitious research units (scientists, research groups, or institutions) with sizes of 50, 100, 500, and 1000 papers. In each case, 1000 samples with a respective size were drawn.

### 3.1 The comparison of the citation impact distributions

Figure 1 shows a comparison of the different approaches (Hazen, InCites, SCImago, and P100) calculated for an exemplary reference set with 50 papers. Since the CWTS approach calculates probabilities of a paper belonging to a pre-specified impact class, e.g., the top 10% most-highly cited papers, instead of a number independent to this pre-specified impact class, it could not be included in this comparison.

Figure 1 shows the comparison of the different approaches with the distribution of the citation counts ("Cites"). All methods lead to a linearization of the skewed citation distribution. The comparison of the approaches of InCites, Hazen, and SCImago shows that they generate similar percentile ranks in the middle and high citation impact area, but not in the low area. This area—the tale of lowly cited papers--obviously makes the difference. The figure also shows that P100 normalizes differently from the other approaches: The lower the citation impact, the greater is the difference between P100 and the three other approaches. As a consequence, the decline of the citation ranks is sharper. This effect is a consequence of the use of the unique citation ranks instead of the ranks of all papers in the reference set for the calculation of the citation impact values: The lower the citation impact, the higher is the probability of tied ranks. This higher probability then leads to lower ranks based on the unique citation distribution compared to percentiles based on the rank of all papers in the reference set.

The distribution for InCite's inverted percentiles shares with P100 that zero-cited papers are not counted for the citation impact. However, this effect is a result of a definition in the case of InCite's ranks and therefore shows an abrupt change in the curve in this case, whereas P100 shows a gradual fading out of the citation impact. Note that the decision to discount zero citations as contributions to the citation impact is consistent with our approach of integrating citation curves as previously argued for in the case of the Integrated Impact Indicator (I3, Leydesdorff & Bornmann, 2011).

**3.2    Analysis of the citation impacts of all papers from 1980 in subsequent years**

Figure 2 shows the number of top 50%, top 10%, top 5%, and top1% papers in years 1 to 31 subsequent to 1980. The papers were attributed to the various rank classes on the base of the ranks which were calculated using the various approaches. One would like to see for all approaches (in an ideal situation) that a constant share of papers is attributed to a specific rank class. A more realistic scenario would be—as the results of Wang (2013) suggested—that especially the first years after publication (in 1980) indicate less reliable attributions. It is well-known in bibliometrics that a short citation window can lead to unreliable prediction of the long-term citation impact of a paper.

The results in Figure 2 show that only the SCImago indicator behaves in the desired way: A relatively constant share of papers is attributed to the top 50%, top 10%, top 5%, and top 1% of papers. This is because SCImago significantly reduces the number of ties by introducing a secondary sort key. Especially the InCites, but also the CWTS and Hazen approaches show discrepancies when using citation impact in the first one to five years after publication as estimators. In these years, (significantly) fewer papers are attributed to the rank classes (especially, the top 50%). The new P100 approach shows numbers completely different from (and lower than) the other approaches. This is not surprising since the unique citation distribution of the reference set was used as a *different* base for assigning citation impact values. The effect of the (large majority of) zero-cited papers, for example, on the percentile attribution is removed in the case of the citation-rank distribution of P100.

The second aspect to our interest is the number of papers which are attributed to the rank classes across the years. Here, one would like to see for each year that about 235,000 papers belong to the top 50% (about half of the total of 475,391 papers from 1980), about 47,000 papers belongs to the top 10%, about 23,000 papers to the top 5%, and about 4,500 papers to the top 1%. In Figure 2 it is clearly visible that the SCImago approach meets these criteria, whereas the other approaches show discrepancies especially in the first one to five years. It seems thus that the consideration of the journal impact (SJR2) as a covariate in the calculation of percentiles provides an advantage for the estimation of the long-term citation impacts (see here Baumgartner & Leydesdorff, in press; Bornmann, Mutz, Marx, Schier, & Daniel, 2011). In the classes top 10%, top 5%, and top 1% it is apparent that the InCites approach outperforms the other approaches (significantly) (Bornmann, in press). This is the effect of the option to optimize using the highest-ranked percentile ranks if papers (that means the journals in which they are published) are assigned to more than a single subject category by Thomson Reuters.

Figure 3 shows the number of top 50%, top 10%, top 5%, and top 1% papers in years 1 to 31 which are in the same class in the last year (31). The results are comparable to the findings in Figure 2. The longer the citation window, the better is the approximation to the long-term citation impact. This holds true for all approaches. Furthermore, the higher the rank class, the worse is the estimation of the long-term impact. In a comparison of the approaches, the InCites approach seems to over-estimate the citation impact. The SCImago approach leads to a different curve in comparison to the other approaches in dependence on the different classes: whereas the percentile ranks are a bit higher in the top 50% class, they are a bit lower in the top 1% class. The additional consideration of the journal impact thus seems to be a disadvantage for the prediction of percentile ranks of the excellent papers.

As shown in Figure 2, the P100 approach leads to citations ranks very different from expected in Figure 3. To make the different approaches comparable, Figure 4 shows the percentages of top 50%, top 10%, top 5%, and top 1% papers in years 1 to 31 which are in the same class in year 31. In this comparison, it is clearly visible that the new approach does not predict total citation impact as well as the other approaches; especially when one focuses on the top 1% class. By using the unique citation distribution as a basis, P100 seems to be more sensitive to annual impact changes than the other approaches.

### 3.3 Analysis of fictitious research units (scientists, research groups, or institutions)

In the second part of the empirical analyses, we are interested in the practical application of the different approaches: How do the (percentile) ranks in the years shortly after publication reflect the long-term citation impact of a research unit (scientists, research groups, or institutions) in year 31? For the same reason as discussed above, (i.e., the CWTS approach gives a set of values depending on a pre-specified percentile rank class), the CWTS approach could not be included in this comparison.

In order to simulate the different size of a unit, we sampled 50, 100, 500, and 1000 papers, and drew 1000 samples in each case (in order to test the reliability of the results). Figure 5 shows Spearman rank-order correlations between the average citation impact value (percentile or citation rank) of a research unit in years 1 to 31 and year 31. As the correlation coefficients in Figure 5 show the results are very similar – independent of the size of the research units (50, 100, 500, or 1000) or the approach: The long-term citation impact of the papers can be reliably estimated only approximately ten years after their publication. Although this is also true for the SCImago approach, this approach performs a bit better in estimating the long-term citation impact in the early years than the other approaches. What are the reasons for the high correlations coefficients starting at this time point which are visible for all approaches? The results of Wang (2013) show that more than half of the cited papers are still going to be cited after ten years, but their received citations are not as many as in the first ten years. Therefore, they do not change their positions any longer in the distribution considerably.

## 4 Discussion

Since the percentile approach has been acknowledged in bibliometrics as a valuable alternative to the normalization of citation counts based on mean citation rates, some different percentile-based approaches have been developed. More recently, two of these approaches have been prominently used in the Leiden Ranking (Waltman, et al., 2012) and SCImago institutions ranking (SCImago Reseach Group, 2012) as evaluation tools. However it is still uncertain which approach should be preferred over other approaches. Furthermore, a solution is still lacking for an unambiguous attribution of percentiles to the discrete citation distributions. The CWTS approach of fractional counting requires a pre-specified citation class and gives results specifically tied to this class, and therefore the approach cannot be used as flexible as the other approaches for further (statistical) analyses.

P100 solves the problem of tied citations and uses a standardized scale (here: from 0 to 100) for the normalization of citation impact in different reference sets. With the introduction of P100 in this study, we have tried to analytically implement these requirements in the specification of citation ranks and rank classes: The ranks scale from 0 to 100 and the problem of tied citations can be solved by using the unique citation distribution for the distinction of ranks.

Although the new approach sounds conceptually clear at first, only empirical analyses can point out its possible advantages when compared with the other approaches. Using a publication set including nearly 500,000 papers published in 1980, we investigated how the different approaches are able to predict the long-term citation impact (in year 31) from citation impacts in previous years (years 1 to 30). In the first analysis, we tested whether the different approaches lead to specific expected values with respect to rank classes (top 50%, top 10%, top 5%, and top 1%) in years 1 to 31.

Since P100 is based on the unique citation distribution, it leads to completely different ranks than the other approaches and is thus hardly comparable. The comparison of the other approaches among themselves reveals that the method used by InCites overestimates citation impact (because of using the highest percentile rank if papers are assigned to more than a single subject category) and the SCImago approach demonstrates surprising capabilities in predicting the long-term citation impact on the basis of citation rates in the first few years. The consideration of journal impact in solving the problem of tied citations seems to have generated this positive effect.

In the second analysis of this study, we tested the ability of the approaches to predict on the basis of their allocation in the various years (1 to 30) the top 50%, top 10%, top 5%, and top 1% papers in the last year (that is, year 31). In this case, all five approaches could be directly compared among each other. The results show a disadvantage in the predictive power of P100 as against the other approaches. These other approaches performed similarly in that they have difficulties to estimate the long-term impact (very) early and these difficulties become greater with a higher rank class.

Since the higher rank classes (e.g., the top 10%) are of special interest in research evaluation studies, bibliometricians should be aware that only a small proportion of high impact papers can be identified within a few years after publication. It is a common practice in bibliometrics to use a citation window of three to five years for measuring citation impact. That means—against the backdrop of our results—that (within three to five years after publication) significantly less than half of the top 1% papers can be identified as top papers in the long run. Similarly alarming results have been already published by Wang (2013) with respect to the indicators normalizing citations on the basis of mean citation rates.

To demonstrate the deficiencies of the percentiles and ranks to estimate the citation impact in the long-run, we simulated the situation where publication sets (e.g., for research evaluation studies) are compiled from WoS. For the simulation, we drew samples with different sizes from the papers published in 1980. Without considering rank classes in this analysis, the results again demonstrate the difficulties of the various approaches to predict long-term citation impact from citations in early years. The SCImago approach performs a bit better in this respect; however, the differences are marginal in this comparison.

For a reliable approximation of long-term citation impact, for example for a correlation coefficient of $r \geq 0.8$, a citation window of at least five years is required. This would mean that five years is the minimum for a citation window in research evaluation studies. As Wang (2013) pointed out further, "there are significant differences in citation ageing between different research fields. For studies on each specific field, a tailored citation window is preferred. For example, if 0.8 is considered as an adequate Spearman correlation for the evaluation, then a 3-year time window may be sufficient for the biomedical research fields and multidisciplinary sciences, while a 7-year time window is required for the humanities and mathematics" (p. 866).

At the beginning of this paper, we questioned the legitimacy of using the arithmetic-average-based citation normalizations, and reviewed recent development in the percentile-based citation normalizations. In this transformation of raw citation counts into citation percentiles, the central issue pertains to the choice of appropriate citation distributions to be normalized, integrated, and evaluated. Mean citation rates are not appropriate for the normalization, because distributions of citations are skewed. In this sense, moving towards distributions of citation percentiles has been a significant improvement. However, there are still normative issues pertaining to the choice of citation distributions not systematically discussed: (1) our new approach highlights two desirable properties, that is, the maximum and minimum citation ranks are 100 and 0 respectively in every reference set. Another desirable property is that (2) the median-rank paper should receive a rank of 50 (which is provided by the Hazen's formula).

(4) Moving from raw citation counts to percentiles, one also loses information in the data. Percentile-based approaches treat the number of citations as an ordinal variable, instead of a cardinal variable, that is, only the order but not the scale is preserved. However, this cardinal-to-ordinal transition may cause problems: For example, in the case of two reference sets, each with 100 papers, one set may contain a paper cited more than 1000 times while all other papers have never been cited, whereas the other set has one paper cited once while all other papers similarly were never cited. In this scenario, percentiles-based approaches would evaluate the papers with 1000 citations similar to the paper with a single citation. However, the author of this heavily cited paper has good reasons to protest against this evaluation.

In summary, a set of desirable properties of the optimal percentile distribution for evaluation can be listed. In addition to properties such as a fixed maximum, a fixed minimum, and the median across reference sets at 50, all cited papers should receive a positive number;

one could try to preserve some aspects of the cardinal nature of the citation counts. Many other desirable properties remain to be specified. However, these properties may be incompatible among them and make it thus impossible to construct a "best" indicator for citations. Therefore, there are still many problems to be solved, and bibliometricians need to (1) be cautious about various approaches of measuring citation impact, (2) be aware of the tradeoffs between different desirable properties, and (3) make choices serving the purpose of their projects. Citation impact is then a metric which requires several selections by bibliometricians.

## 5   Conclusions

Three important conclusions can be drawn from the present study:

1. The citation window issue confirms conclusions previously drawn by Wang (2013) and Baumgartner and Leydesdorff (in press). This conclusion is detrimental to using short-term citation indicators such as the Journal Impact Factor based on two years (Garfield, 1972) and SNIP based on three years (Waltman, van Eck, van Leeuwen, & Visser, 2013) for the prediction of the citation impact of papers. Indicators based on short citation windows measure only "transient" impact at the research fronts and not "sticky knowledge claims".

2. The findings of this study show that the consideration of journal impact improves the prediction of long-term impact. This leads to questions for further research: Is it possible to improve citation impact measurements on the base of short citation windows by the consideration of covariates, such as journal impact, the number of authors, and other possible factors influencing citation counts (Bornmann & Daniel, 2008).

3. P100 underperforms empirically although it is conceptually convincing. Thus, the question is: Are there changes possible in the specification of P100 which would improve the empirical performance of this indicator while keeping the conceptual advantages?

# Acknowledgements

The data used in this paper are from a bibliometrics database developed and maintained by the Competence Center for Bibliometrics for the German Science System (KB) and derived from the 1980 to 2011 Science Citation Index Expanded (SCI-E), Social Sciences Citation Index (SSCI), Arts and Humanities Citation Index (AHCI), Conference Proceedings Citation Index- Science (CPCI-S), and Conference Proceedings Citation Index- Social Science & Humanities (CPCI-SSH) prepared by Thomson Reuters (Scientific) Inc. (TR®), Philadelphia, Pennsylvania, USA: ©Copyright Thomson Reuters (Scientific) 2012. The authors thank the KB team for its collective effort in the development of the KB database. We are grateful to Ronald Rousseau for exchanges of ideas.

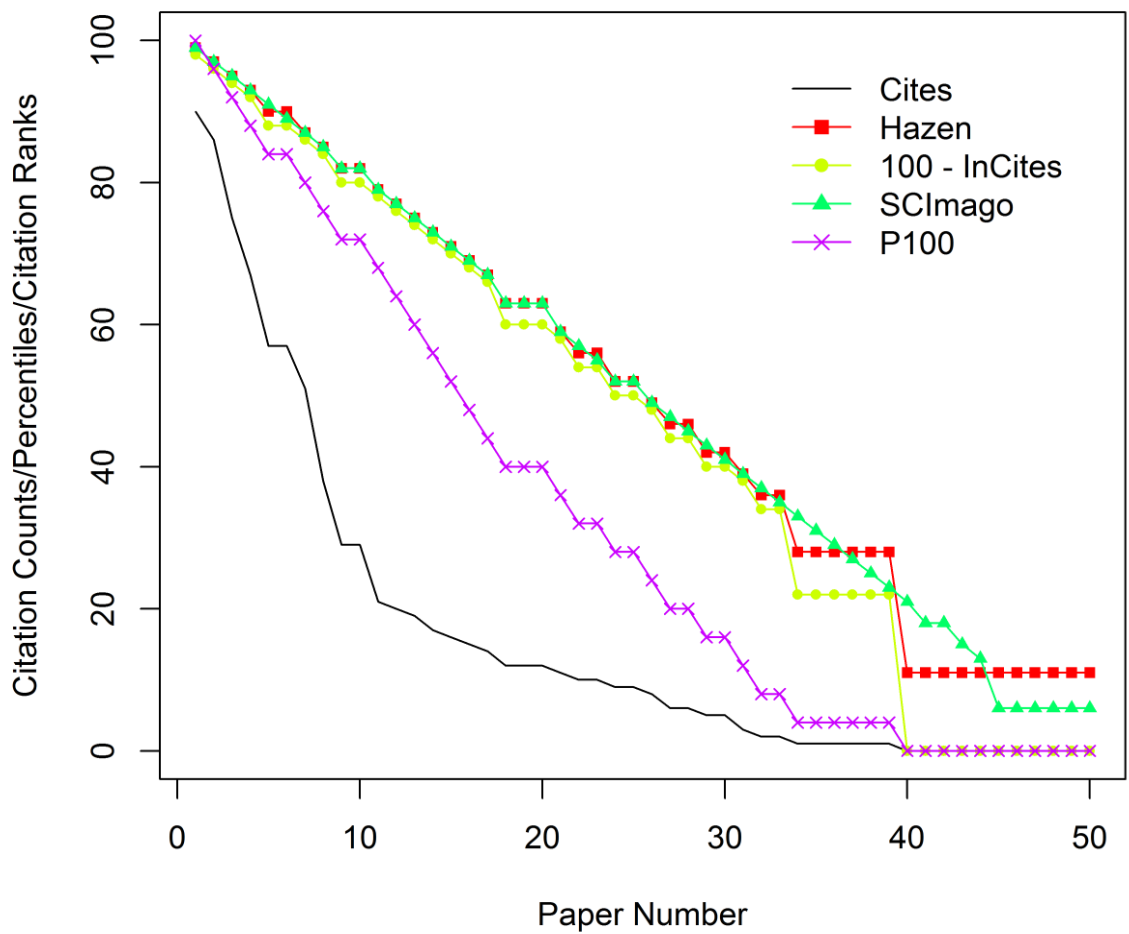

Figure 1. Comparison of different (percentile-based) approaches calculated for a reference set with 50 papers. The X-axis ("Paper Number") is the rank of papers, and papers are sorted in descending order of their citations. The Y-axis ("Citation Counts/Percentiles/Citation Ranks") is the raw number of citation counts for *Cites*, the value of citation percentiles for the percentile-based approaches, and the value of citation ranks for P100.

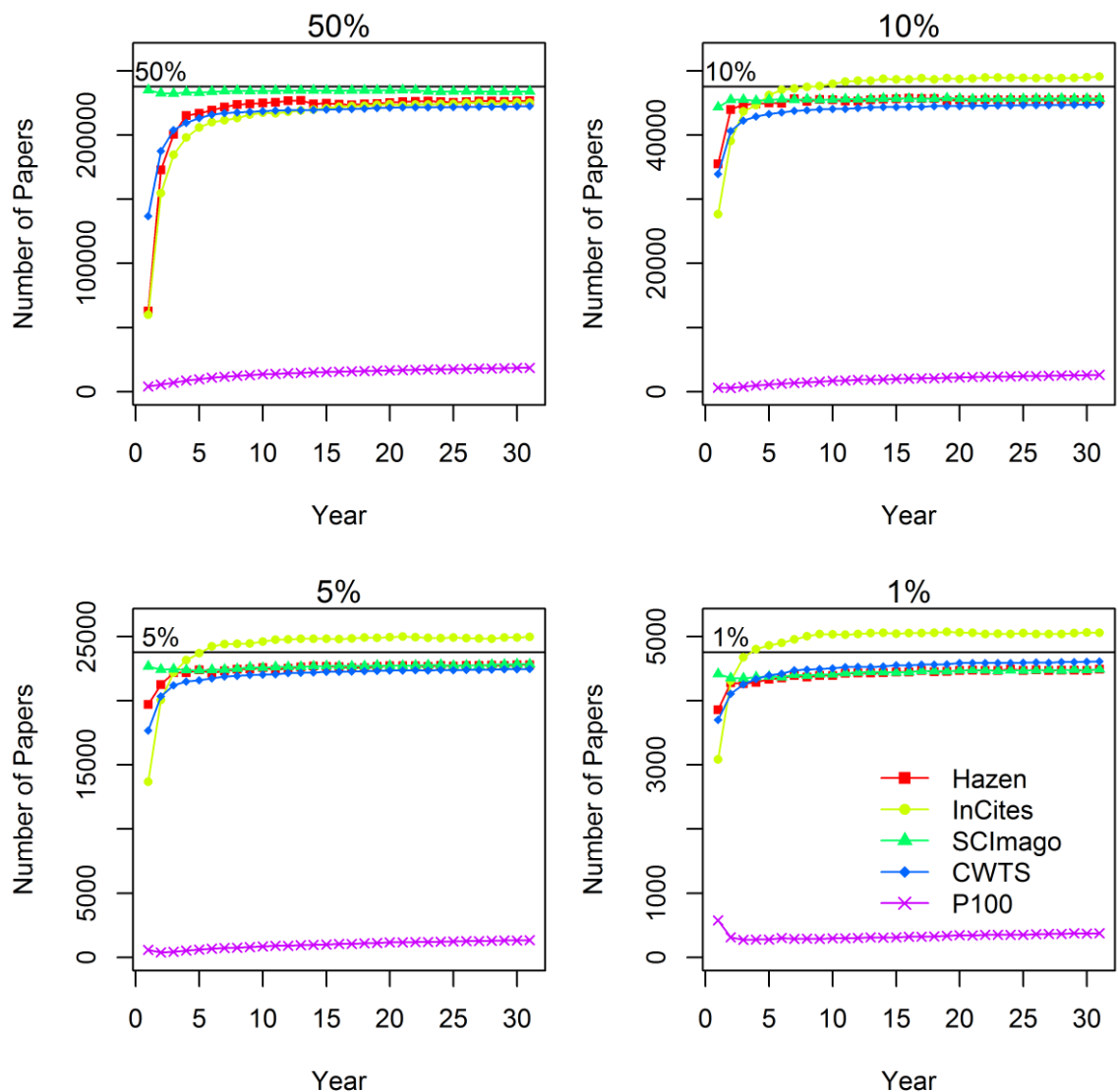

Figure 2. Number of top 50%, top 10%, top 5%, and top1% papers in year *t* (database: 475,391 papers from 1980). X-axes ("Year") index *t*, and *t* is the year after publication (i.e. year 1–31 correspond to 1980–2010 respectively).

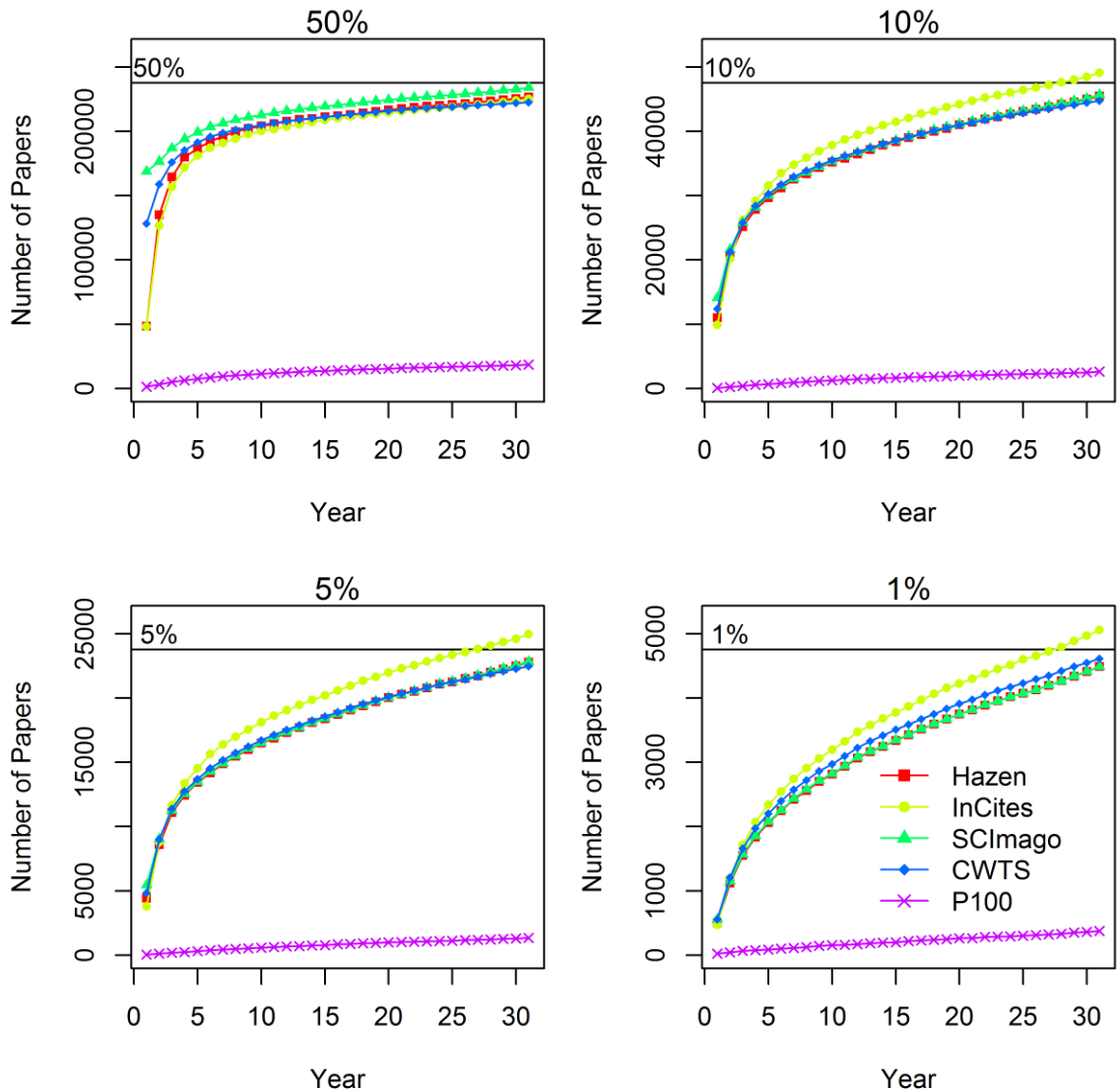

Figure 3. Number of top 50%, top 10%, top 5%, and top 1% papers in year *t* which are in the same class in year 31 (database: 475,391 papers from 1980).

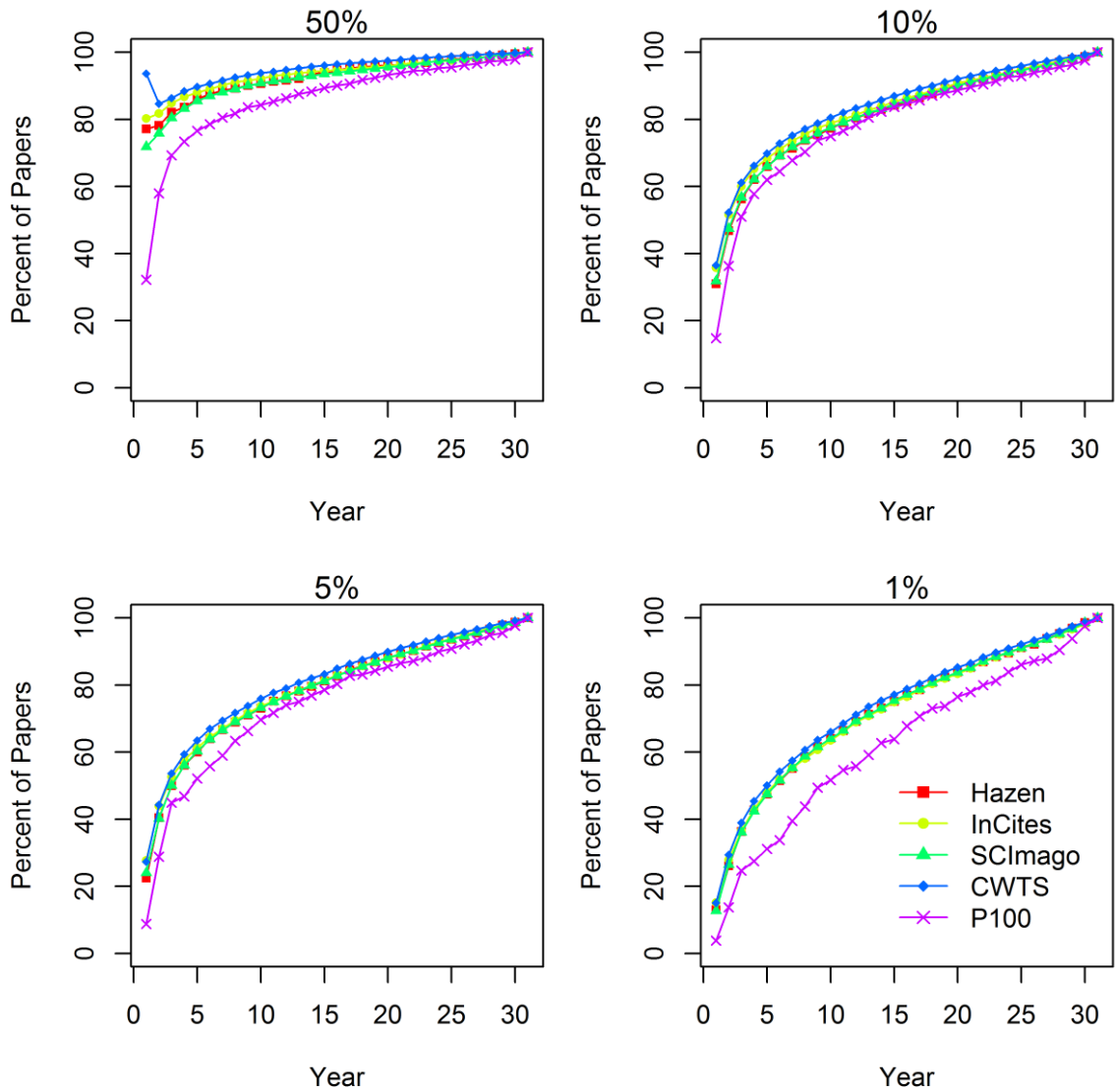

Figure 4. Percent of top 50%, top 10%, top 5%, and top 1% papers in year *t* which are in the same class in year 31 (database: 475,391 papers from 1980).

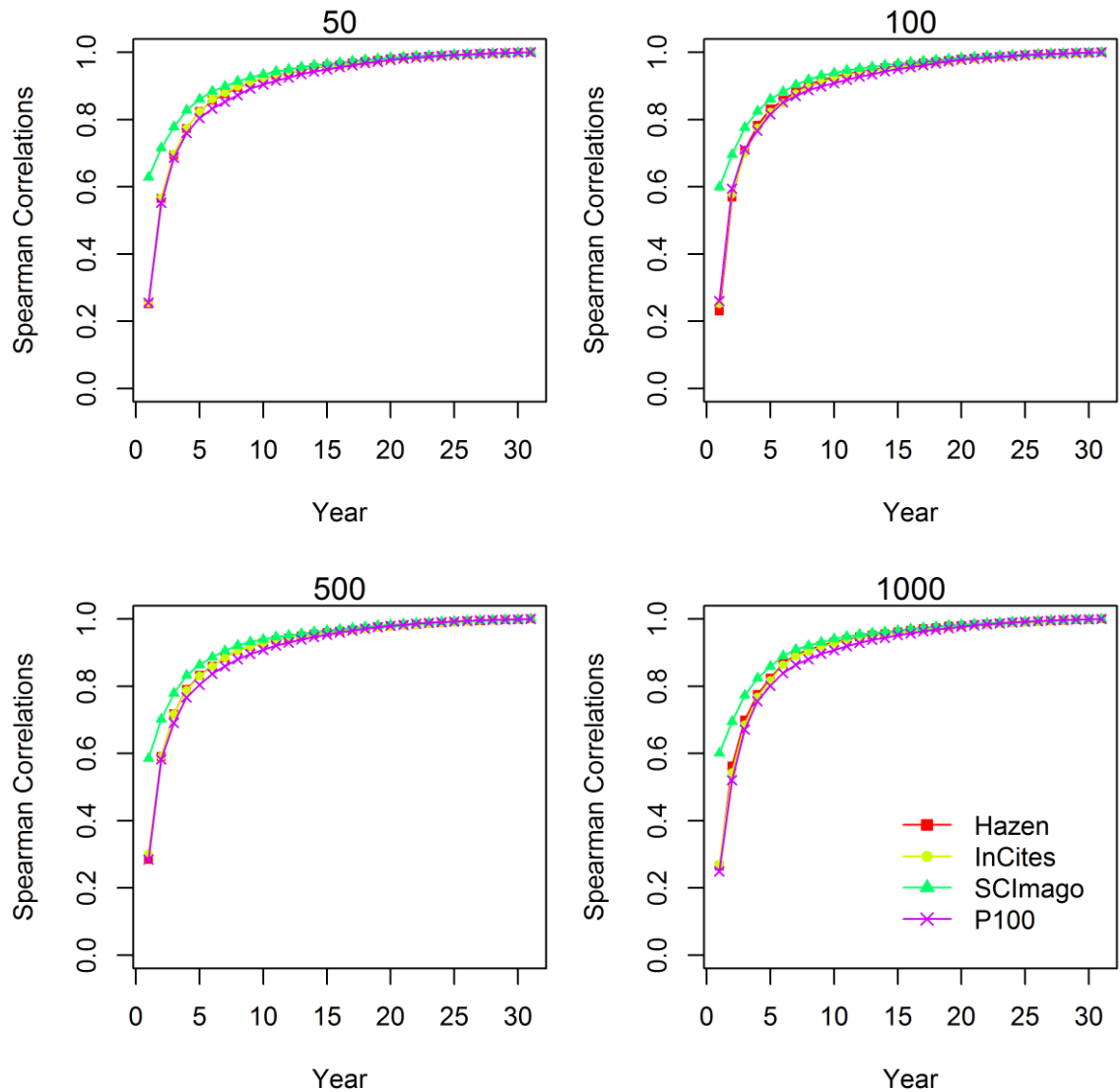

Figure 5. Spearman rank-order correlations between the average percentile or citation rank of a research unit in years *t* and year 31 (random samples of 1000 research units with 50, 100, 500, and 1000 papers drawn from the papers published in 1980).

Table 1. A first fictitious example (without considering papers in the reference set having the same citation counts; the order in the table is inverted)

| Citations | Rank i | Citation rank | Rank for the paper under study |
|---|---|---|---|
| 10 | 6 | 100 | |
| 7 | 5 | 83 | |
| 4 | 4 | 66.4 | 66.4 |
| 3 | 3 | 49.8 | |
| 2 | 2 | 33.2 | |
| 1 | 1 | 16.6 | |
| 0 | 0 | 0 | 0 |

Table 2. A second fictitious example (with papers in the reference set having the same citation counts; the order in the table is inverted)

| Papers with citations | Unique citation values | Citation rank | Rank for the paper under study |
|---|---|---|---|
| 130 | 130 | 100 | |
| 90 | 90 | 85.8 | |
| 90 | | | |
| 90 | | | |
| 90 | | | |
| 40 | 40 | 71.5 | |
| 38 | 38 | 57.2 | 57.2 |
| 32 | 32 | 42.9 | |
| 32 | | | |
| 32 | | | |
| 7 | 7 | 28.6 | |
| 4 | 4 | 14.3 | |
| 4 | | | |
| 4 | | | 14.3 |
| 0 | 0 | 0 | |
| 0 | | | |
| 0 | | | |
| 0 | | | |